# Imaging real-time amorphization of hybrid perovskite solar cells under electrical biasing

*Min-cheol Kim[1][†], Namyoung Ahn[2][†], Diyi Cheng[3][†], Mingjie Xu[4,5], Xiaoqing Pan[4,5,6], Suk Jun Kim[7], Yanqi Luo[1], David P. Fenning[1,3,8], Darren H. S. Tan[1], Minghao Zhang[1], So-Yeon Ham[3], Kiwan Jeong[2], Mansoo Choi[2,9]\*, Ying Shirley Meng[1,3,8]\**

[1]Department of NanoEngineering, University of California San Diego, 9500 Gilman Drive, La Jolla, CA 92093, USA

[2]Global Frontier Center for Multiscale Energy Systems, Seoul National University, Seoul 08826, Republic of Korea

[3]Materials Science and Engineering Program, University of California San Diego, 9500 Gilman Drive, La Jolla, CA 92093, USA

[4]Irvine Materials Research Institute, University of California Irvine, Irvine, CA, 92697, USA

[5]Department of Materials Science and Engineering, University of California Irvine, Irvine, CA 92697, USA

[6]Department of Physics and Astronomy, University of California Irvine, Irvine, CA 92697, USA

[7]School of Energy, Materials and Chemical Engineering, Korea University of Technology and Education, Cheonan 31253, Republic of Korea

[8]Sustainable Power & Energy Center (SPEC), University of California San Diego, La Jolla, CA 92093, USA

[9]Department of Mechanical Engineering, Seoul National University, Seoul 08826, Republic of Korea

Correspondence and request for materials should be addressed to Y.S.M and M.C.

(email: smeng@ucsd.edu, mchoi@snu.ac.kr)

[†] These authors contributed equally to this work

**ABSTRACT**

Perovskite solar cells have drawn much attention in recent years, owing to its world-record setting photovoltaic performances. Despite its promising use in tandem applications and flexible devices, its practicality is still limited by its structural instability often arising from ion migration and defect formation. While it is generally understood that ion instability is a primary cause for degradation, there is still a lack of direct evidence of structural transformation at the atomistic scale. Such an understanding is crucial to evaluate and pin-point how such instabilities are induced relative to external perturbations such as illumination or electrical bias with time, allowing researchers to devise effective strategies to mitigate them. Here, we designed an *in-situ* TEM setup to enable real-time observation of amorphization in double cation mixed perovskite materials under electrical biasing at 1 V. It is found that amorphization occurs along the (001) and (002) planes, which represents the observation of *in-situ* facet-dependent amorphization of a perovskite crystal. To reverse the degradation, the samples were heated at 50 °C and was found to recrystallize, effectively regaining its performance losses. This work is vital toward understanding fundamental ion-migration phenomena and address instability challenges of perovskite optoelectronics.

**INTRODUCTION**

Since the onset of the 21$^{st}$ century, increasing demands for clean energy generation to displace traditional fossil fuels have generated an influx of interest in solar, wind and other renewable sources of energy. Of these various technologies, photovoltaics are highly promising to achieve sustainable, safe and robust energy to meet society's growing needs, by tapping only a fraction of the huge amounts of solar energy entering the earth's atmosphere. Today, solar generation market products are still dominated by conventional silicon-based technologies. However, a combination of poor energy conversion efficiencies and low throughput still motivate the exploration of better alternatives to serve both traditional solar markets, as well as emerging applications in tandem or flexible devices.[1,2] As such, organic-inorganic hybrid perovskite solar cells (PSCs) have been extensively investigated in last decade due to their high power conversion efficiencies (PCE), exceeding 25.5% in recent reports,[3] as well as lower fabrication costs.[4] While initial reports indicate short device lifetime, they were mainly attributed to the material's sensitivity to heat, moisture or other external factors which has been largely addressed with strategies such as introduction of more hydrophobic compounds or improved cell level designs with encapsulation.[5,6] However, the community soon realized that another significant cause of poor lifetime arises from material intrinsic degradation as well, believed to be a result of crystal phase instability and generation of defects. This instability is attributed to the fact that the ionic crystal exhibits excessive ion migration under light illumination and electrical bias.[7,8] Given that the application of any solar cells demand long term operation under wide weather and climatic conditions, strategies to either mitigate or reverse these degradative effects are vital.

In PSCs, ion vacancies are easily generated since they have relatively low vacancy formation energies.[9,10] Under a potential bias, ions and vacancies migrate easily toward electrodes induced by coulombic forces while the device is illuminated.[11-13] Namely, the charged ions (or carriers) will be simply governed by drift-diffusion model. As a consequence, ion accumulation and charge carrier accumulation at the respective interfaces leads to performance degradation within just a few hours.[14,15] Fortunately, these devices with degraded performance can recover their initial capabilities if the ions or accumulated charges can be redistributed, which typically occurs after being stored in a dark state.[16,17] However, when this process is repeated over many cycles under periodic electrical bias, it has been found that the device eventually experiences irreversible degradation, preventing its practical use in long term devices.[15,18]

These degradation effects have been investigated to some extent in the literature and reported as a consequence of ion migration and segregation. Diego *et al.* recently reported the integrated intensity of main peaks using X-ray diffraction (XRD) patterns of PSCs and found decreases in intensities when the sample was biased at 1.2 V in the dark, concluding that perovskite amorphization occurred as a result of ion migration.[19] While these observations provide a clue on the amorphization effects under electrical bias, the structural transformation at atomic-levels still remains unclear.[20-22] Alternatively, element-tracking techniques were previously employed to reveal that organic cations and halide anions migrate during device operation.[14,23,24] These experiments have collectively concluded that facile movements of halide anions impair the performance of PSCs, agreeing with past theoretical predictions using density functional theory calculations.[7,9,25] Although such interpretations can explain the reversible nature of ion redistribution, crystallographic transformation by ion migration is still not verified. Direct observation of ion migration at the atomistic scale is needed to provide evidence for ion migration-induced amorphization. To this end, high resolution transmission electron microscopy (TEM) imaging can potentially provide the missing knowledge and help gain fundamental understanding on such structural changes, phase transitions and amorphization. Unfortunately, this is experimentally challenging to execute, due to the dynamic nature of perovskite materials instability against application of external electrical bias, making it hard to capture without simultaneous electrical bias and imaging.

In this study, we addressed this limitation by designing an *in-situ* TEM setup using a micro e-chip that can apply a stable electrical bias while imaging the bias-induced sample simultaneously. This allows real-time observation of amorphization within the double cation mixed perovskite materials under a 1 V forward bias. It was observed that amorphization occurs in the preferred orientation corresponding to the (001) and (002) planes, which is the first reported observation of *in-situ* facet-dependent amorphization of a perovskite crystal. These results were supported with X-ray diffraction (XRD) measurements, that confirmed the partial amorphization from reduced (001) and (002) peak intensities, while the (111) peak was maintained after bias. Additionally, the effect of the partial amorphization was examined via a light-induced degradation test; a forward bias induced by photovoltage led to faster performance decay compared to a controlled sample without electrical bias (short-circuited). To reverse these effects, mild heat treatment of the degraded sample at 50 °C was used to drive recrystallization of the amorphous phase, allowing it to regain its performance. Based on these observations, we propose a possible mechanism of the partial amorphization. This work demonstrates the ability to achieve real-time imaging of perovskite amorphization and provides an effective method to address the ion instability challenges of PSCs.

## RESULTS & DISCUSSION

**In situ observation of amorphization induced by electrical bias.** To study nanoscale changes of perovskite materials under electrical bias, we employed *in-situ* TEM using a micro e-chip setup, where a stable electrical bias was applied to a nano solar cell TEM lamella. (**Fig. 1a**) As shown in **Fig. 1b**, a nano solar lamella was prepared using a focused ion beam (FIB) methodology developed from Lee et al.[26] The lamella was then mounted between the electrodes of an e-chip. In order to apply an electrical bias through the perovskite layer without shorting the cell, a column was cut at the top-right corner and bottom-left corner of the lamella. The middle region (5 μm in length) was thinned down to around 120 nm in thickness to facilitate TEM observation. Detailed fabrication procedure of the nano solar cell and electrical connection verification can be found in **Supplementary Fig. 1 & 2.** Note that the voltage bias applied in this experiment was set to mimic the photovoltage of open-circuited PSC device under one sun illumination. The sample was prepared at the pristine state and was not pre-biased before the measurement.

First, the effect of a forward bias (1 V for 5 min) on the cross-sectional morphology was examined, where no morphological change or damage were observed, indicating the stability of nano solar cell under such beam condition. (Details included in **Supplementary Video 1** and **Supplementary Fig. 3**). In the higher magnification, obvious bias-induced changes were presented in several domains of perovskite nanocrystals as can be seen in **Supplementary Video 2** and **Supplementary Fig. 4.** To obtain the nanoscale structural information, high-resolution TEM (HRTEM) images were acquired as shown in **Fig. 1c** and **Supplementary Video 3**. Lattice fringes of perovskite crystals at 0 s were seen to gradually disappear (highlighted region in yellow boxes of **Fig. 1c**), and became completely absent after 300 s of continuous 1 V forward bias. Real-time fading of specific lattice fringes by electrical bias implied amorphization of perovskite nanocrystals, and such structural changes were not reported before in organic-inorganic hybrid perovskite materials at atomic scales. A further comparison between the HRTEM images at 0 s and 300 s was shown in **Supplementary Fig. 5**, where reduced fast Fourier transform (FFT) patterns were sampled at different regions. Interestingly, the lattice fringes in the yellow boxes disappeared after 5 min of electrical bias, however, those in the red box have retained majority of its crystalline characteristics post-bias. A detailed analysis showed that the d-spacings of the lattice fringes in the yellow boxes and red boxes were corresponding to the (001) plane of perovskite crystal ($(FAPbI_3)_x(MAPbBr_3)_{1-}$

$_x$) (6.36 Å)[27] and (001) plane of PbI$_2$ (6.96 Å),[28] indicating that only a partial amorphization has occurred.

It is noted that such observations might be a result of sample tilt or beam damage induced by continuous beam irradiation. As such, selected area electron diffraction (SAED) patterns were acquired to provide essential information to correlate the lattice fringe disappearance with the partial amorphization of the perovskite materials. **Fig. 2** shows the SEAD patterns at two different regions (Region A and B from **Fig. 2c**) to compare the effects of electrical bias. At region A, SAED patterns were acquired under continuous electron beam irradiation without applying electrical bias, where no significant change after 5 mins of beam irradiation was observed (**Figs. 2a and 2b**). Thus, it is deduced that amorphization is not induced by beam damage. Conversely, changes in SAED patterns were noticed at the region B where 1 V of electrical bias was applied for 5 mins in the absence of electron beam irradiation (**Figs. 2d and 2e**). The amorphous ring became obviously more pronounced in the SAED pattern of region B, which is in good agreement with the HRTEM images discussed earlier (**Fig. 1**), confirming the amorphization effects of electrical bias. Collected SAED patterns are indexed from crystal structure data of perovskite and compared to simulated electron diffraction pattern (**Supplementary Fig. 6**).[27] The amorphous ring mostly overlapped with $d$-spacing of (002) diffraction spots of perovskite in the SAED patterns, which is a strong evidence of the partial amorphization. The magnified SAED patterns at the region B (**Fig. 2f**) clearly revealed that the amorphous ring corresponding to (002) diffraction spot was becoming more apparent. Curiously, the diffraction spots of [111] and [022] groups remained unchanged during the entire amorphization process of the [002] group. This finding suggested that the amorphization of perovskite material occurs under a facet-dependency, which has been reported for other polycrystalline materials in the literature.[29] Such partial amorphization might originate from the most active halide species migration rather than cation species migration, maintaining the organic cation cage and PbI$_6$ octahedral but losing the electron diffraction intensity along (002) direction from the iodide evacuation.

Since observations made in the *in-situ* TEM can be attributed to localized phenomena, bulk scale characterization was also conducted to confirm the partial amorphization at the device level. *Ex-situ* XRD measurements of the fresh and post-bias device (1 V for 10 min) were conducted and shown in **Fig. 3**. Magnified XRD patterns in **Fig. 3b and c** showed that the forward bias led to a significant decrease in peak intensities of (001) and (002) planes while peak intensities of PbI$_2$ and FTO remained the same, confirming that amorphization only occurred in (001) and (002) planes of the perovskite materials. The change in peak intensity of (111) plane was not significant as compared to (001) and (002) planes, which was consistent with the TEM results showing the facet-dependent amorphization (**Fig. 3d**). During bias-induced amorphization, phase transition to $\delta$-phase did not occur as shown in **Supplementary Fig. 7**, which means the structural change under electrical bias is not from either decomposition or phase transition but from the amorphiztion.[30] These *in-situ* TEM and *ex-situ* XRD results imply that partial amorphization of the perovskite film is induced by electrical forward bias, consistent with previous studies.[19,21] Additionally, this affirms the correlation between atomic structural changes and the performance degradation of photovoltaic devices under the electrical bias.

**Amorphization effects on performance.** To address PSC long term stability issues, it is essential to understand how such partial amorphization affects photovoltaic performance. To evaluate this, a degradation test setup was used to apply different electrical conditions; open-circuit (OC) and short-circuit (SC), as illustrated in **Figs. 4a** and **d** to explore the effect of amorphization on the PSC's performance. When both PSCs were placed under continuous one sun illumination, the photovoltage (~ 1 V) was developed in the OC device and charge carriers were accumulated at the interfaces between perovskite and transport layers, creating an photovoltage induced electric field for charged ions pointing from anode to cathode.[31,32] On the other hand, charge carriers were not accumulated at the interfaces in the SC device, thereby no induced electrical potential between cathode and anode was applied to charged ions. In other words, photovoltage induced electric field from the charge accumulation was only generated in OC condition which was equivalent to the forward bias to the device, analogous to that during the *in-situ* TEM and *ex-situ* XRD measurements. Considering the injection of carriers from either external electrical bias or light illumination allowed the ions to migrate more freely, applying the electrical bias or using the induced electric field from the charge accumulation could give the same structural changes caused by ion migration.[33] The degradation test was conducted in pure nitrogen environment in order to exclude the effect of surrounding reactive molecules as oxygen and moisture. **Figs. 4b** and **e** showed changes in *J-V* curves of open-circuited and short-circuited devices, respectively. Photovoltaic parameters calculated from each *J-V* curve were listed in **Table S1**. For both conditions, performance degraded when the device was illuminated and restored after recovering in dark due to photo-activated reversible ion migration.[17] However, the forward-bias in the OC case caused dramatic changes during degradation as compared to the SC case, consistent with previous studies.[16,17,34,35] Specifically, the open-circuit voltage ($V_{oc}$) largely decreased after 16 h of light soaking and did not return to the original value after recovery in darkness, which indicates that the defect concentration and the crystallinity of the perovskite film were mainly influenced by forward-bias (see **Fig. 4c**).[36,37] On the other hand, the $V_{oc}$ values hardly changed in the SC case and recovered to the initial $V_{oc}$ after 16 h of recovery as shown in **Fig. 4f**. Unlike $V_{oc}$ values, degraded power conversion efficiency (PCE) of perovskite devices in both OC and SC case recovered to the initial PCE as shown in **Supplementary Fig. 8**. The partially amorphous phase may be re-crystallized spontaneously while the device is stored in dark under inert atmosphere and ambient temperature, which can possibly lead to performance recovery. To exclude the possibility of phase segregation being involved during device degradation and recovery process, we measured the photoluminescence (PL) spectra of perovskite device before and after the light soaking at OC, and mapped the spatial variation of PL intensity and peak center position respectively (see **Supplementary Fig. 9**). Peak shift or splitting were not observed while PL intensity increased, which confirms the absence of phase segregation during the experiments. Charge accumulation is mainly responsible for increase of PL intensity, which clearly show the formation of photovoltage under light illumination.[38,39]

**Thermally induced performance recovery**. Considering that degradation as a result of ion migration can be reversed with ion redistribution under darkness,[14,19] supplying additional thermal energy should promote such ion redistribution and hasten the recovery process. Thus, a thermal recovery test was conducted on the degraded devices by heating it at 50 °C in darkness (**Fig. 4g**). The degraded PSC under continuous light illumination was restored to the initial performance within just 3 h when they were heated at 50 °C in darkness (**Figs. 4h and i**). Note that performance was not fully recovered even after 10 h of storage in dark at room temperature, demonstrating the effectiveness of applying mild heating. These results indicate that thermal energy facilitated the prompt recovery process of degraded PSCs through recrystallization of amorphous perovskite. Although a reversible change of PCEs of PSCs has been reported from previous studies[14,16,17,40-42], this work offers new direct observation of dynamic changes in crystallinity of perovskite materials as a result of forward bias. Moreover, mild heating was found to be an effective solution to expedite performance recovery under darkness, potentially extending the practical lifetime of PSC devices if they can be recovered at increased temperatures periodically during nighttime.

**Electrochemical effects of ion migration induced amorphization.** To elucidate electrochemical effects of partial amorphization, electrical bias-induced ion migration inside the perovskite device was investigated via electrochemical impedance spectroscopy (EIS) and energy-dispersive X-ray spectroscopy (EDS) measurements. After applying various magnitudes of forward bias, inductance response (or negative capacitance) was observed at low frequencies for bias greater than 1 V as shown in **Fig. 5**. Since the device was measured and stored in dark, there would be no influence from photogenerated charges and generation of electrical potential. While the device was biased, the electrical circuit of the device was merely composed of resistances and capacitances similar to general perovskite devices (**Figs. 5a-c**).[43] Thus, the new inductance component after forward bias beyond 1 V would be indicative of a significant material transformation in terms of ionic movement and migration (**Figs. 5d and e**).[44-47] The ionic transformation was possibly due to the partial amorphization based on our *in-situ* TEM observations (see **Fig. 2**).

There are two possible scenarios for the inductance response; 1) the ion-channeling effect as a consequence of amorphization or, 2) generation of negative capacitance. Ion-channeling leads to the formation of current channels that can generate inductance component in EIS, as non-uniform current densities flowing through the perovskite layer can generate magnetic flux within the device, finally forming the induced electromotive force. For instance, resistance-switching random access memory devices employing a metal oxide layer also exhibits such an inductance component when current channels (ionic nanofilament) are opened by electrical bias.[48] The other possibility of negative capacitance[49], could be a result of iodine ions evacuated from perovskite crystals during amorphization and migrating to interfaces via the electric field and finally accumulated to form a surface voltage. Charge carriers were strongly affected by the surface voltage, from which the interface acted as the negative capacitance when the partial amorphization happened. This was likely related with the inverted hysteresis behavior of the aged device where forward scan (scan from $J_{sc}$ to $V_{oc}$) PCE became larger than the reverse scan one (scan from $V_{oc}$ to $J_{sc}$) (see **Supplementary Fig. 10**). Perovskite device that initially exhibited normal hysteresis (the reverse scan PCE is larger than the forward scan one) showed inverted hysteresis after 16 h of aging under light illumination and OC condition. Since the hysteresis behavior of PSCs was attributed to the ion migration,[8,50] the change of the hysteresis behavior was related with the mobile ions verified by the inductance developed under the forward bias over 1 V.

From the above characterization and analysis results, a possible mechanism for the partial amorphization coupled with dynamics of ion migration is proposed. Although some defects with low defect formation energies are intrinsically formed in pristine perovskite crystals, this still represents a low fraction of defect density (**Fig. 5f**). After an external electric field is applied via forward bias or illumination, the formation of Frenkel pairs ($V_I^+/I_i^-$) will be more energetically favorable due to their low defect formation energy (**Fig. 5g**).[7] The newly-formed interstitial defects and vacancies are the most active species during electric field-induced ion migration.[13] Our EDS results shows that iodine distribution exhibits the most drastic change whereas other components of perovskite crystals remained relatively unchanged after the forward bias (**Supplementary Fig. 11**). Moreover, iodine accumulation near the cathode interface was observed as shown in **Supplementary Fig. 12**, suggesting that negatively charged interstitial iodine defects migrated under the driving force of the electric field. The resulting atomic collisions of iodine migration would possibly induce the partial amorphization observed (**Fig. 5h**). This mechanism of cascading atomic collisions along with the local accumulation of defect concentration have been proposed in previous studies as well, further supporting the amorphization mechanism discussed here.[51,52] Additionally, facet-dependent amorphization can be explained by the observations that [001] and [002] groups (which contain a larger number of iodides compared to [111] groups) likely experienced iodide loss due to electrical bias. Consequently, amorphization would be more favorable with the increased defect concentrations in the [001] and [002] groups. Finally, these degradation effects can be reversed by resting the device under dark conditions, which allows recrystallization and ion redistribution to occur (**Fig. 5i**). To increase the rate of recovery, the process can be facilitated by adding mild thermal energy as shown in the results (**Fig. 4i**).

# CONCLUSIONS

In conclusion, *in-situ* high-resolution TEM was used to achieve direct real-time observation of perovskite crystal amorphization. A lamella of a nano solar cell was prepared using FIB and mounted on an e-chip to observe structural changes under a constant forward bias of 1 V. Under bias, the disappearance of lattice fringes in HRTEM images accompanying with manifestation of a clear amorphous ring in SAED patterns was observed, which is evidence of partial amorphization. These bias-induced partial amorphization results were complemented with bulk XRD measurements, which showed a decrease of the peak intensities along the (001) plane, while other peaks remained unchanged. A possible mechanism of partial amorphization based on halide ion migration and self-implantation is proposed. As ion migration was identified as the primary cause of performance degradation, performance recovery rate was found to improve by heating the PSC device at 50 °C to promote ion re-distribution, where the device was fully recovered after only 3 hours. Overall, this study provides unprecedented observations of atomic-scale transitions during bias-induced degradation. These observations reveal that although partial amorphization results in degradation in PSC devices, it can be effectively recovered through re-crystallization process under mild conditions, reflecting a realistic strategy to maintain long term device performance.

# METHODS

***Perovskite device fabrication.*** All the chemical solvents for perovskite precursor and methylammonium chloride (MACl) were purchased from Sigma-Aldrich and used as received. Lead iodide ($PbI_2$) and $SnO_2$ nanoparticle dispersion were purchased from Alfa Aesar. Formamidinium iodide (FAI) and methylammonium bromide (MABr) were purchased from Greatcellsolar. Pre-etched indium doped tin oxide (ITO)/glass substrate was cleaned by sonication immersed in detergent, deionized (DI) water, acetone and isopropanol (IPA) for 10 min respectively. Cleaned ITO/glass substrate was surface-treated by air-plasma cleaner for 10 min to improve wettability. $SnO_2$ thin film was deposited on the substrate by spin-coating at 4,000 rpm for 30 s using $SnO_2$ nanoparticle dispersion (tin(IV) oxide, 15 % in $H_2O$ colloidal dispersion) further diluted in DI water (2.67 wt%). The $SnO_2$ thin layer was annealed at 150 °C for 30 min. Mixed cation and halide PSCs (($FAPbI_3$)$_x$($MAPbBr_3$)$_{1-x}$) was prepared by 2-step sequential spin-coating procedure inside of $N_2$-filled glovebox.[53] 1.5 M of $PbI_2$ in the mixture of dimethylformamide (DMF) and dimethyl sulfoxide (DMSO) (DMF:DMSO = 0.9:0.1) was prepared for $PbI_2$ layer deposition and a 1-mL solution of FAI (90 mg), MABr (9 mg), and MACl (9 mg) in IPA were prepared for mixed organic solution. $PbI_2$ solution was spin-coated onto the $SnO_2$ layer at 2,300 rpm for 30 s, then annealed at 75 °C for 1 min to form $PbI_2$ layer. Mixed organic halide solution was then distributed on the semi-transparent $PbI_2$ film, then spin-coated at 2,500 rpm for 30 s. The reacted perovskite layer was annealed at 150 °C for 10 min outside the glovebox. Spiro-MeOTAD layer was deposited on the perovskite layer by dynamic spin-coating at 2,300 rpm for 30 s using Spiro-MeOTAD solution (50 mg of Spiro-MeOTAD in 550 μL of chlorobenzene, mixed with 19.5 μL of 4-tert-Butylpyridine, 11.5 μL of Li-TFSI solution (540 mg /1 mL of acetonitrile), and 5 μL of FK 209 Co(III) TFSI solution (375 mg / 1 mL of acetonitrile). A layer of Au metal (30 nm) was then deposited on top of the Spiro-MeOTAD with the rate of 0.2 Å/s over a metal shadow mask to form the metal electrode under high vacuum at $10^{-7}$ torr.

***Nano-solar cell for in situ TEM specimen preparation.*** The surface and cross-section morphology of the perovskite solar cell sample was observed by FIB/SEM (Scios DualBeam, FEI), which was then used to prepare the nano solar cell for in situ TEM observation. The operating voltage of electron beam was 5 kV. Emission current of electron beam was set to 50 pA to minimize potential beam damage on perovskite layer during fabrication. An argon ion beam source was used to mill and thin the sample. The operating voltage of ion beam source was 30 kV. Different emission currents of ion beam were chosen for different purposes, i.e. 10 pA for imaging by ion beam, 0.1 nA for cross-section cleaning/lamella thinning and 3 nA for pattern milling. After conventional FIB liftout process, the lamella was mounted on an electrical biasing E-chip (E-FEF01-A4, Protochips) and connected to the gold electrodes with FIB-deposited organometallic Pt to secure the electrical conduction pathways. When the lamella was thinned down to 300 nm, it was cut free at the anode region in the top (Au) layer and cathode region in the bottom (ITO) region to prevent the nano solar cell from shorting and enable electrical biasing. Final thickness of the lamella is ~120 nm after thinning to be electron transparent. In situ biasing in FIB chamber followed a similar fabrication protocol and experimental setup described in a previous work.[26]

***In situ TEM measurement.*** The *in situ* TEM measurement under electrical bias was performed at Irvine materials research institute (IMRI) with a Protochips Fusion Select holder on a JEOL 2800 TEM operating at 200 keV. The *in situ* TEM specimen was mounted on a FIB-optimized E-chipTM (E-FEF01-A4, Protochips). Ultra-low current detectable potentiostat (SP-200, Biologic) was utilized to monitor the current in real time and to apply the electrical bias simultaneously. Images are collected with a low beam current at spot size 5 to reduce electron beam damage. *In situ* TEM video was captured at 4k resolution and 25fps (Oneview IS, Gatan). SAED patterns were collected with the second smallest aperture and the region was carefully selected to avoid undesired cross-damage from the electron beam.

***Characterization of perovskite solar cells.*** The current-voltage characteristics were measured by temperature-controlled I-V tester with Class A solar simulator (Newport) under AM 1.5G at 100 mW cm$^{-2}$ at room temperature with the scan speed of 0.1 V/s for both reverse and forward scan. The light intensity was calibrated by using a Si reference cell before the measurement. The aperture size was 0.07 cm$^2$.

***Degradation and recovery test.*** PSCs were aged under solar simulator (Newport, AM 1.5G at 100 mW cm$^{-2}$ calibrated) inside the glovebox filled with pure Nitrogen gas at OC and SC condition respectively. Source meter (2400, Keithley) was disconnected to imitate the OC condition, while source meter was connected and photocurrent were detected during the SC condition. Aged PSCs were stored in the dark box which block the light to perform dark recovery process. Aged PSCs were stored on the hotplate (50 °C) surrounded by the dark box to perform thermal recovery process. Initial, aged and recovered performance of PSCs were characterized by measuring the current-voltage characteristics and 8 different devices were measured for both recovery processes.

***Electrochemical impedance spectroscopy.*** All EIS measurements were conducted in a dark chamber using an electrochemical workstation (Autolab320N, Metrohm, Switzerland). We applied AC voltage with different magnitudes (from 0.6 V to 1.1 V in steps of 0.1 V) and a fixed amplitude of 10 mV. The frequency of applied AC voltage varied from 1 mHz to 1 MHz. ZView software (Scribner Associates, USA) was used to fit impedance spectra.

***Ex situ XRD measurement.*** Ex situ XRD analysis of electrical bias-induced changes of perovskite materials was collected using a Rigaku SmartLab X-ray diffractometer with Cu Kα source operating at 30 kV and 15 mA with a step size of 0.01º at 0.9º/min, scanning over 10-60º. PSCs were applied with 1 V of forward electrical bias for 10 min to induce bias-induced amorphization. PSCs before and after applying electrical bias were rinsed with Chlorobenzene to remove Spiro-MeOTAD and Au electrode.

***Micro-photoluminescence (μ-PL) mapping.*** A Renishaw inVia confocal Raman microscope equipped with a Si CCD detector and a motorized sample stage was utilized to conduct the μ-PL mapping at 20× magnification. A 633-nm incident laser with 5.68 μW power and 0.01 s exposure time was scanned in 20 μm steps. The incident photons interacted with the absorber by penetrating through the glass/ITO and the electron transport layer (ETL, $SnO_2$), and the emission luminescence was collected via a 1200 mm grating.

**ACKNOWLEDGEMENTS**

This work was supported by California Energy Commission EPIC Advance Breakthrough Award (EPC-16-050). This work was also supported by the Basic Science Research Program through the National Research Foundation of Korea (NRF) funded by the Ministry of Education (2019R1A6A3A12031412) and Global Frontier R&D Program of the Center for Multiscale Energy Systems (2012M3A6A7054855). The authors would like to acknowledge the financial support received for this study from the U.S. Department of Energy, Office of Basic Energy Sciences, under Award Number DE-SC0002357 (program manager Dr. Jane Zhu). The work done at the University of California Irvine was supported by the Irvine Materials Research Institute (IMRI).

**AUTHOR INFORMATION**

Affiliations

**Department of NanoEngineering, University of California San Diego, La Jolla, CA, USA**

Min-cheol Kim, Yanqi Luo, David Fenning, Darren H. S. Tan, Minghao Zhang & Ying Shirley Meng

**Global Frontier Center for Multiscale Energy Systems, Seoul National University, Seoul, Republic of Korea**

Namyoung Ahn, Kiwan Jeong & Mansoo Choi

**Materials Science and Engineering Program, University of California San Diego, La Jolla, CA, USA**

Diyi Cheng, So-Yeon Ham & Ying Shirley Meng

**Irvine Materials Research Institute, University of California Irvine, Irvine, CA, USA**

Mingjie Xu & Xiaoqing Pan

**Department of Materials Science and Engineering, University of California Irvine, Irvine, CA, USA**

Mingjie Xu & Xiaoqing Pan

**Department of Physics and Astronomy, University of California Irvine, Irvine, CA, USA**

Xiaoqing Pan

**School of Energy, Materials and Chemical Engineering, Korea University of Technology and Education, Cheonan, Republic of Korea**

Suk Jun Kim

**Sustainable Power & Energy Center (SPEC), University of California San Diego, La Jolla, CA, USA**

David Fenning & Ying Shirley Meng

**Department of Mechanical Engineering, Seoul National University, Seoul, Republic of Korea**

Mansoo Choi

CONTRIBUTIONS

M.K., N.A., M.C. and Y.S.M. conceived and designed the study. M.K. and S.H. fabricated and characterized the perovskite photovoltaic devices. D.C. prepared the in situ TEM specimens, and M.X. and X.P. performed the in situ TEM analysis and M.X., X.P., M.K., D.C., M.Z. and S.J.K. processed and analyzed the in situ TEM data. M.K. and D.C. performed the ex situ XRD measurement. M.K., N.A. and K.J. performed degradation and recovery test for the device, and N.A. and K.J. performed impedance spectroscopy analysis. Y.L. and D.F. performed the photoluminescence mapping, and all authors reviewed and extensively discussed the results. M.K., N.A., D.C., D.H.S.T, M.C. and Y.S.M. wrote the manuscript with input from all authors. M.C. and Y.S.M. led the research project.

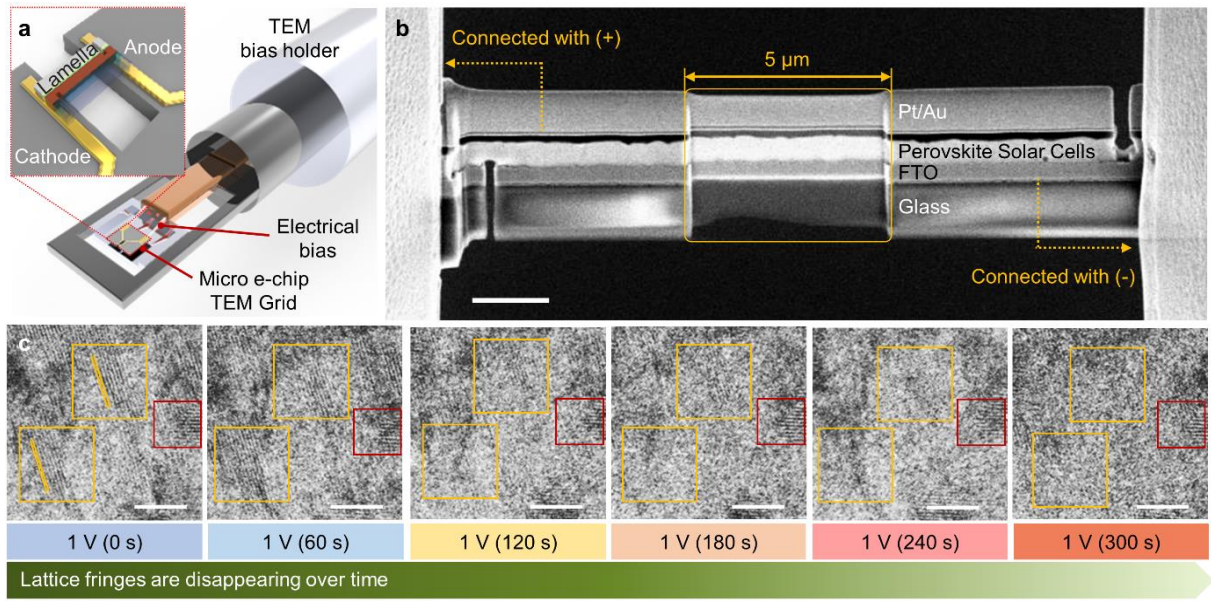

**Figure 1. In situ observation of perovskite materials under electrical bias.** (a) Schematic of in situ TEM sample configuration under electrical bias. A nano solar cell lamella is mounted on the micro e-chip TEM and stimulated with electrical bias in a TEM biasing holder. (b) SEM image of the nano solar cell lamella used for in situ observation prepared by FIB. The scale bar is 2 μm. (c) HRTEM image series of in situ observation on perovskite materials under 1 V electrical bias. Lattice fringes of perovskite highlighted in yellow boxes are gradually disappearing while the lead iodide lattice fringes in the red box are retained during the 5-min biasing. The scale bars are 10 nm.

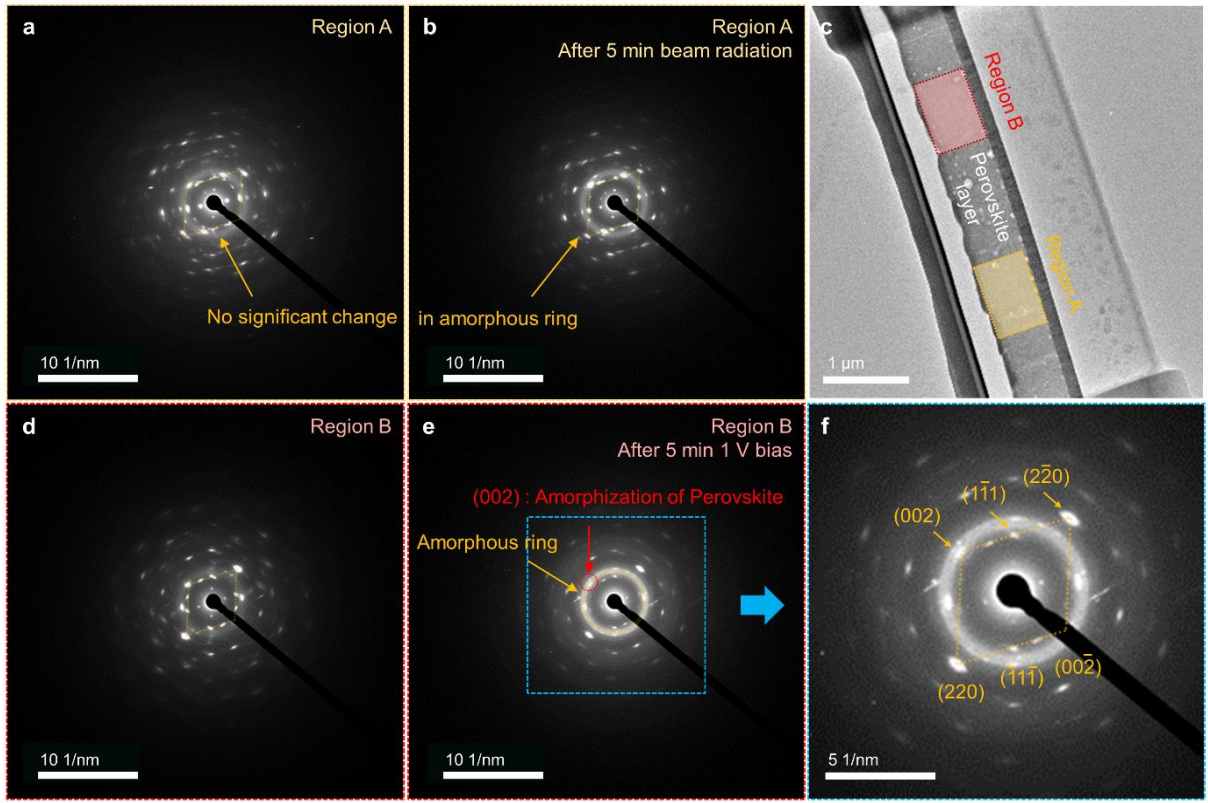

**Figure 2. Verification of perovskite amorphization induced from electrical bias instead of beam damage.** SAED patterns of perovskite materials at region A (a) before and (b) after 5-min continuous beam radiation without electrical bias. (c) TEM image of nano solar cell indicating the locations of region A and region B. SAED patterns of perovskite materials at region B (d) before and (e) after 5-min forward bias (1 V) without beam radiation. (f) Zoomed-in SAED patterns of blue region in (e) with main diffraction spots indexed.

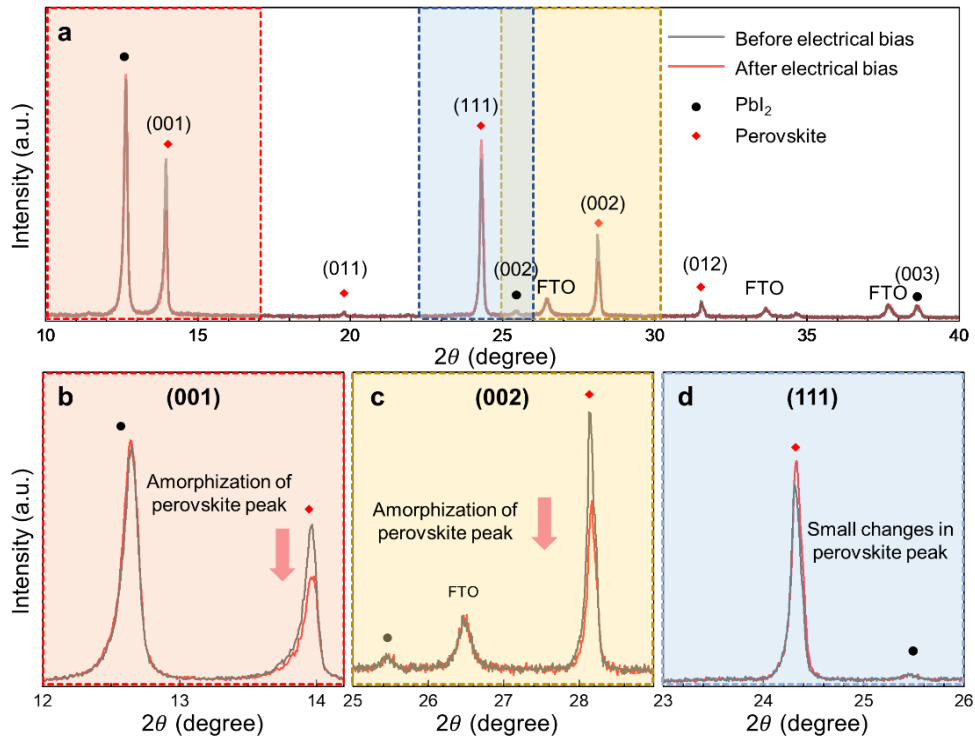

**Figure 3. Perovskite film amorphization under electrical bias verified from XRD.** (a) XRD patterns of perovskite materials at pristine state (black line) and after applying forward bias (1 V) for 10 min (red line). The measured perovskite solar cells are rinsed by chlorobenzene to remove Spiro-MeOTAD and Au layer. Zoomed-in XRD patterns highlighting the differences near (b) (001), (c) (002) and (d) (111) peaks of perovskite materials, where $PbI_2$ and perovskite species are indicated by black dots and red diamonds respectively.

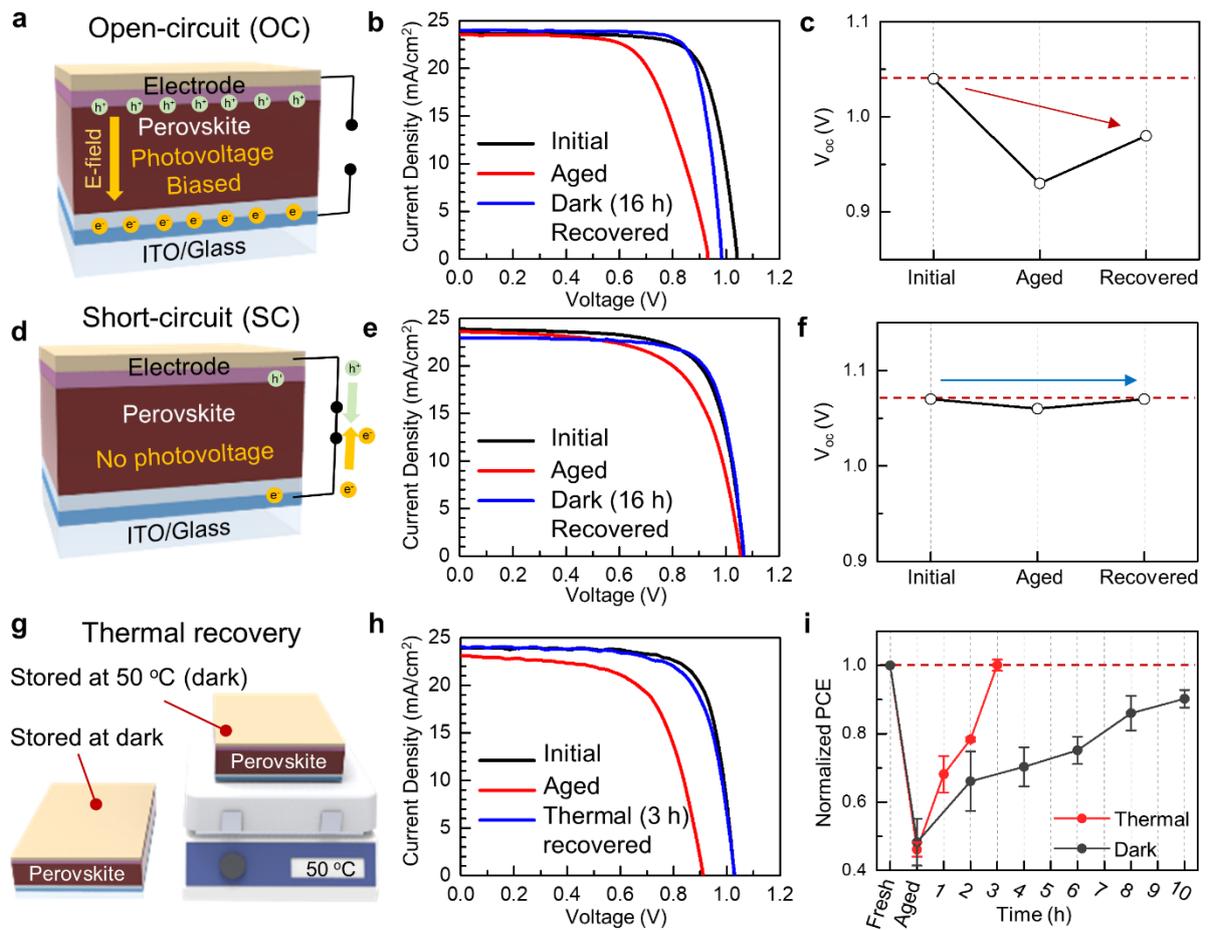

**Figure 4. Device performance degradation under electrical bias and dark/thermal recovery.** (a, d) Schematics of perovskite solar cells under light illumination at open-circuit (OC) and short-circuit (SC) conditions. The photovoltage generated at OC condition can induce charge accumulation that leads to the formation of an internal electric field at the perovskite layer. (b, e) Current density ($J$) – Voltage ($V$) curves of perovskite solar cells at initial state (black line), after 16 h of aging (red line), and after 16 h of recovery at dark (blue line) at OC and SC conditions, respectively. (c, f) Evolution of open circuit voltage ($V_{oc}$) from the initial state to recovered state for perovskite solar cells at OC and SC conditions. (g) Schematic of dark recovery and thermal recovery of degraded perovskite solar cells. (h) *J-V* curves of perovskite solar cells at initial state (black line), after 16 h of aging at OC condition (red line), and after 3 h of recovery at 50 °C. (i) Comparison of thermal and dark recovery of degraded perovskite solar cells.

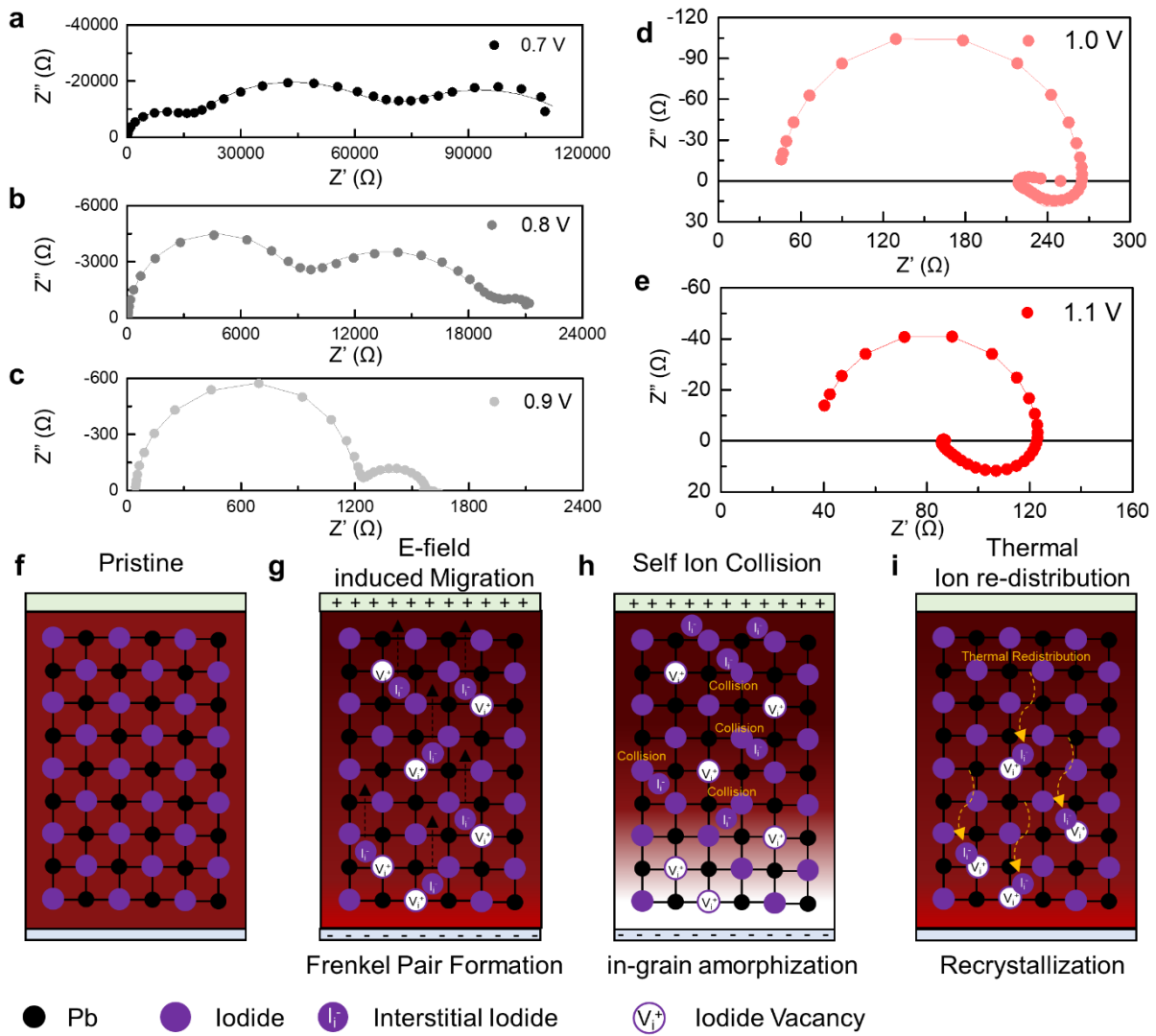

**Figure 5 Evidence for ion migration and mechanism of amorphization from the ion migration.** EIS plots of the perovskite solar cells measured from 1 MHz to 100 mHz with forward bias of (a) 0.7 V (b) 0.8 V (c) 0.9 V (d) 1.0 V and (e) 1.1 V at dark condition. Inductance (or negative capacitance) occurred only when the forward bias exceeds 1.0 V, verifying the ion migration. (f-i) Proposed mechanism for ion migration-induced partial amorphization of perovskite materials and the recovery from thermal ion redistribution.

# Supplementary Information

# Imaging real-time amorphization of hybrid perovskite solar cells under electrical bias

*Min-cheol Kim[1][†], Namyoung Ahn[2][†], Diyi Cheng[3][†], Mingjie Xu[4,5], Xiaoqing Pan[4,5,6], Suk Jun Kim[7], Yanqi Luo[1], David P. Fenning[1,3,8], Darren H. S. Tan[1], Minghao Zhang[1], So-Yeon Ham[3], Kiwan Jeong[2], Mansoo Choi[2,9]\*, Ying Shirley Meng[1,3,8]\**

[1]Department of NanoEngineering, University of California San Diego, 9500 Gilman Drive, La Jolla, CA 92093, USA

[2]Global Frontier Center for Multiscale Energy Systems, Seoul National University, Seoul 08826, Republic of Korea

[3]Materials Science and Engineering Program, University of California San Diego, 9500 Gilman Drive, La Jolla, CA 92093, USA

[4]Irvine Materials Research Institute, University of California Irvine, Irvine, CA, 92697, USA

[5]Department of Materials Science and Engineering, University of California Irvine, Irvine, CA 92697, USA

[6]School of Energy, Materials and Chemical Engineering, Korea University of Technology and Education, Cheonan 31253, Republic of Korea

[7]Sustainable Power & Energy Center (SPEC), University of California San Diego, La Jolla, CA 92093, USA

[8]Department of Mechanical Engineering, Seoul National University, Seoul 08826, Republic of Korea

Correspondence and request for materials should be addressed to Y.S.M. and M.C.

(email: smeng@ucsd.edu, mchoi@snu.ac.kr)

[†] These authors contributed equally to this work

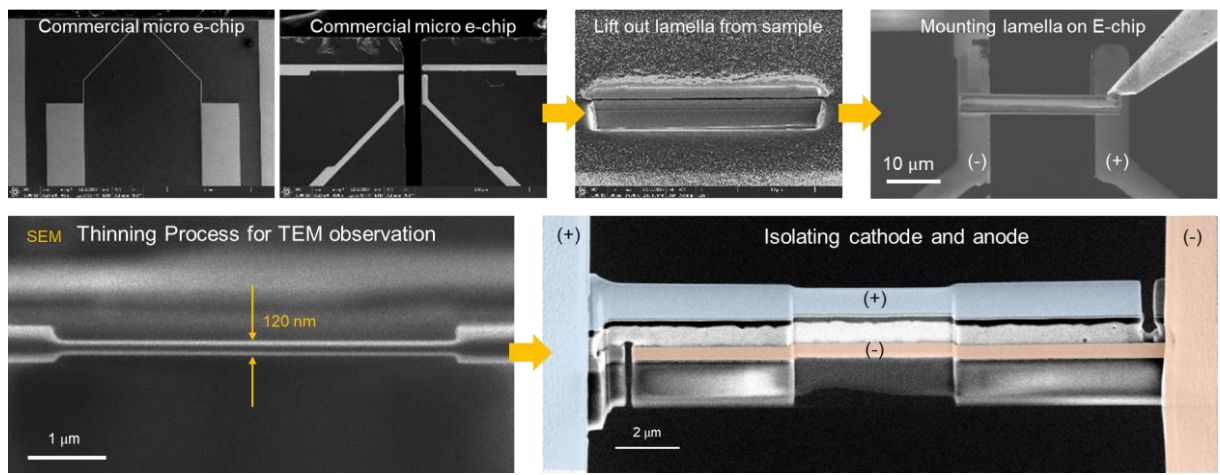

**Fig. S1. Fabrication procedure of microchip-based nano solar cell**

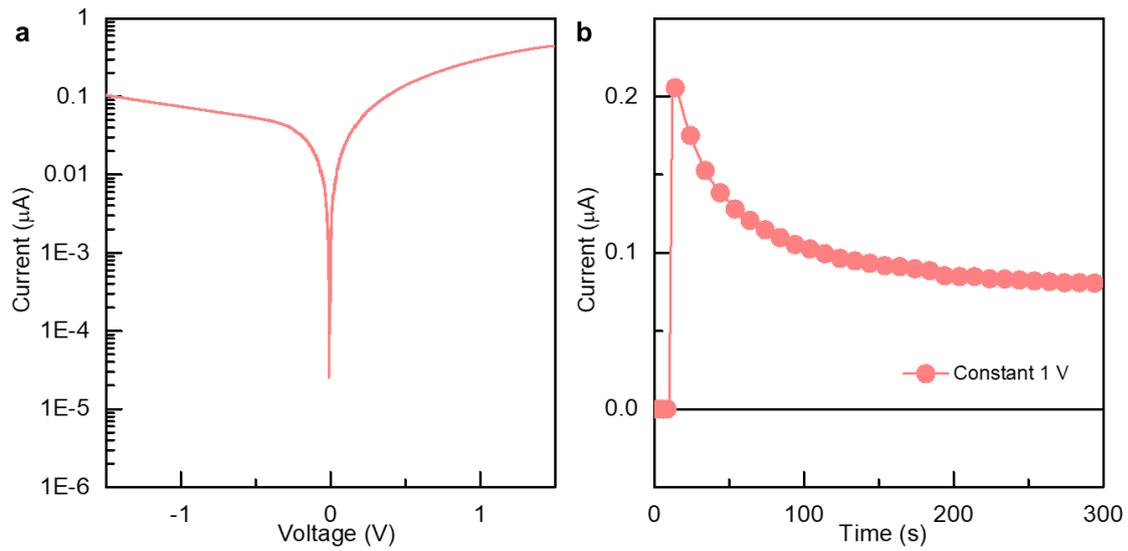

**Fig. S2. JV curves of nano solar cell tested in TEM column and the current response at in situ TEM measurement** (a) Current-voltage response of nano solar cell lamella mounted on the micro-chip prepared by FIB. (b) Current – time plot of nano solar cell lamella measured under constant voltage of 1 V.

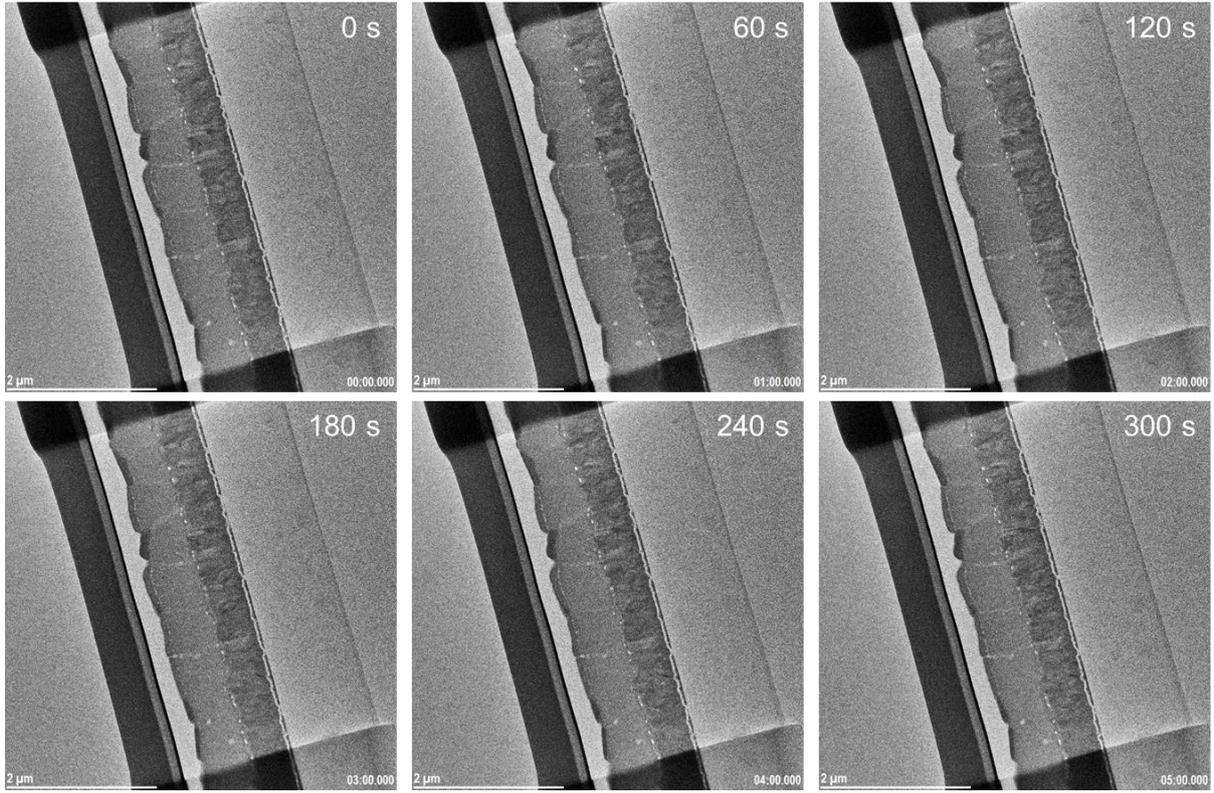

**Fig. S3. In situ TEM images for nano solar cell under electrical bias of 1 V for 5 min.** Sequential cross-sectional TEM images under electrical bias for 5 min.

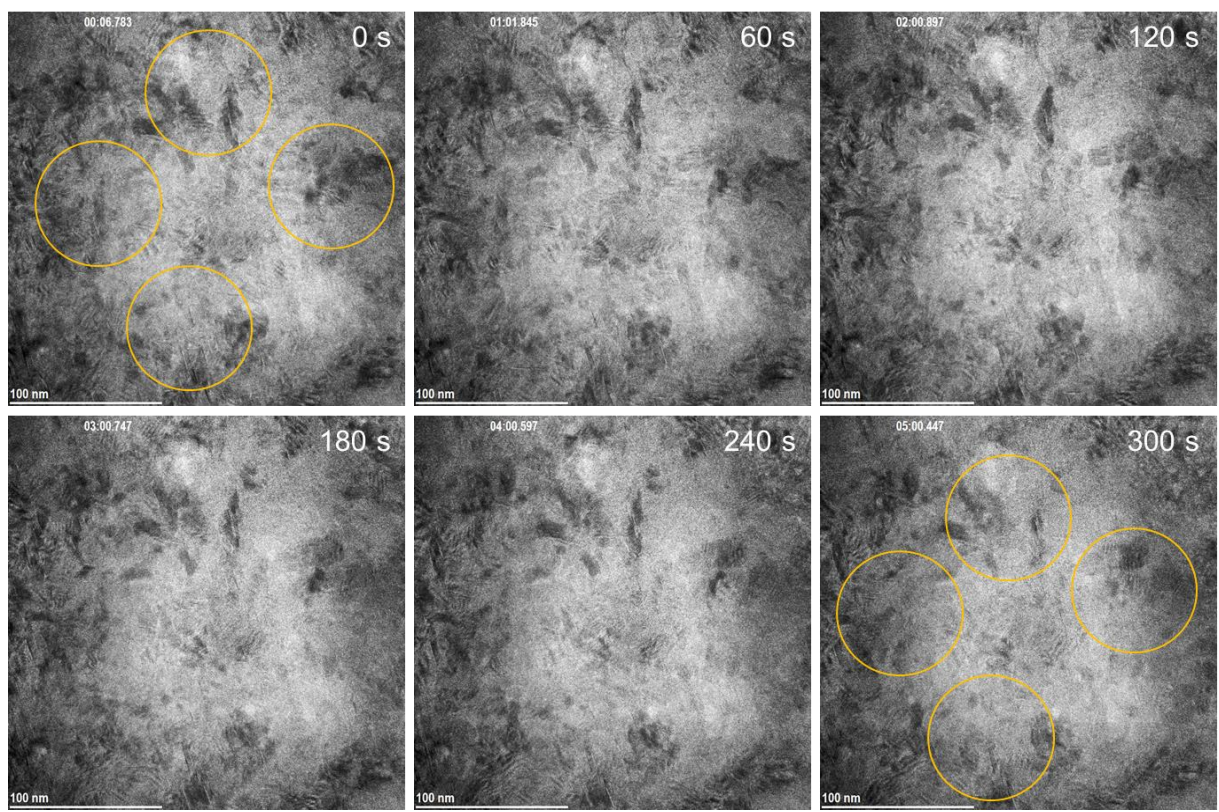

**Fig. S4. In situ HRTEM images for nano solar cell under electrical bias of 1 V for 5 min.**
Sequential cross-sectional HRTEM images under electrical bias for 5 min.

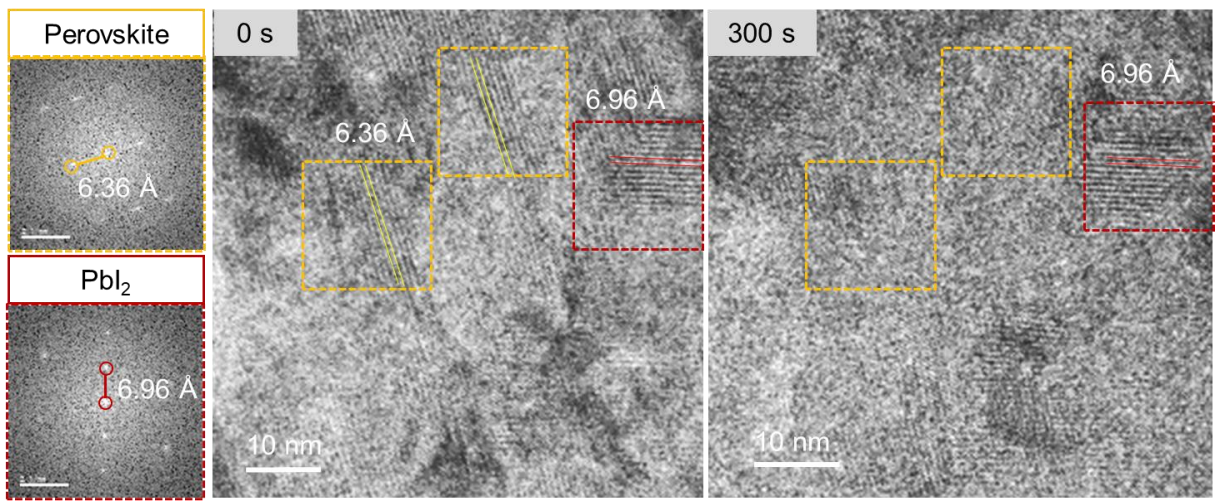

**Fig. S5. In situ HRTEM images and FFT patterns for perovskite materials (yellow box) and PbI$_2$ (red box) at 0 s and 300 s under electrical bias.** A d-spacing for the yellow box indicates the perovskite materials, while that for the red box indicates the PbI$_2$ materials. It is obvious that perovskite materials are partially amorphized as a disappearance of the lattice fringes over applying the electrical bias.

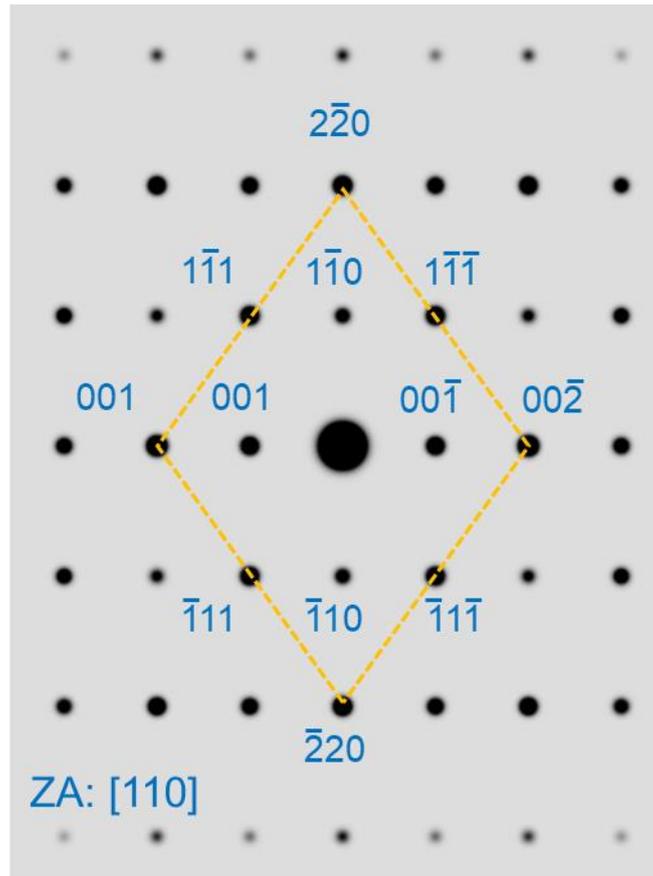

**Fig. S6.** Simulated electron diffraction pattern of perovskite materials (FA$_x$MA$_{(1-x)}$PbI$_3$) with the [110] zone axis by SingleCrystal software.

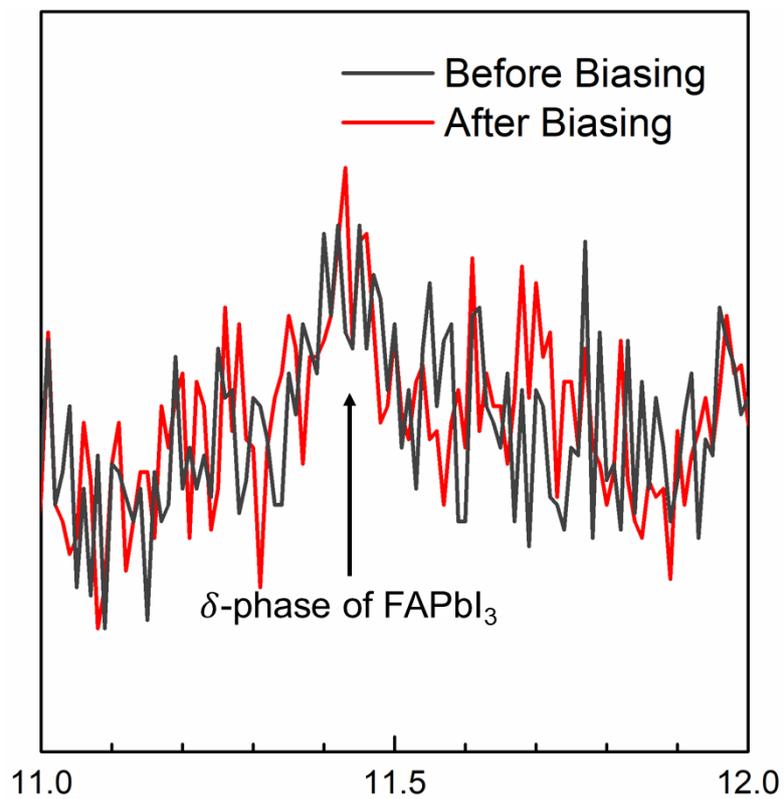

**Fig. S7.** Zoomed in XRD pattern in the range of 11-12 2-theta degrees of perovskite films before and after applying 1 V of electrical bias.

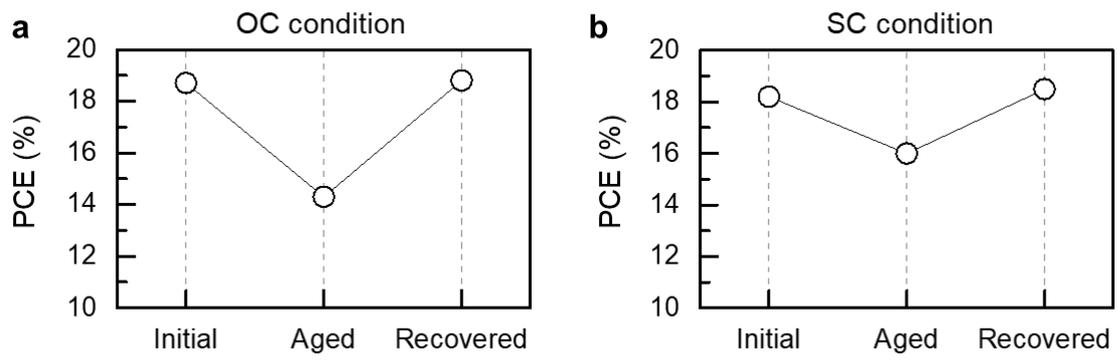

**Fig. S8.** Evolution of Power conversion efficiency (PCE) from the initial state to recovered state for perovskite solar cells at (a) OC and (b) SC conditions.

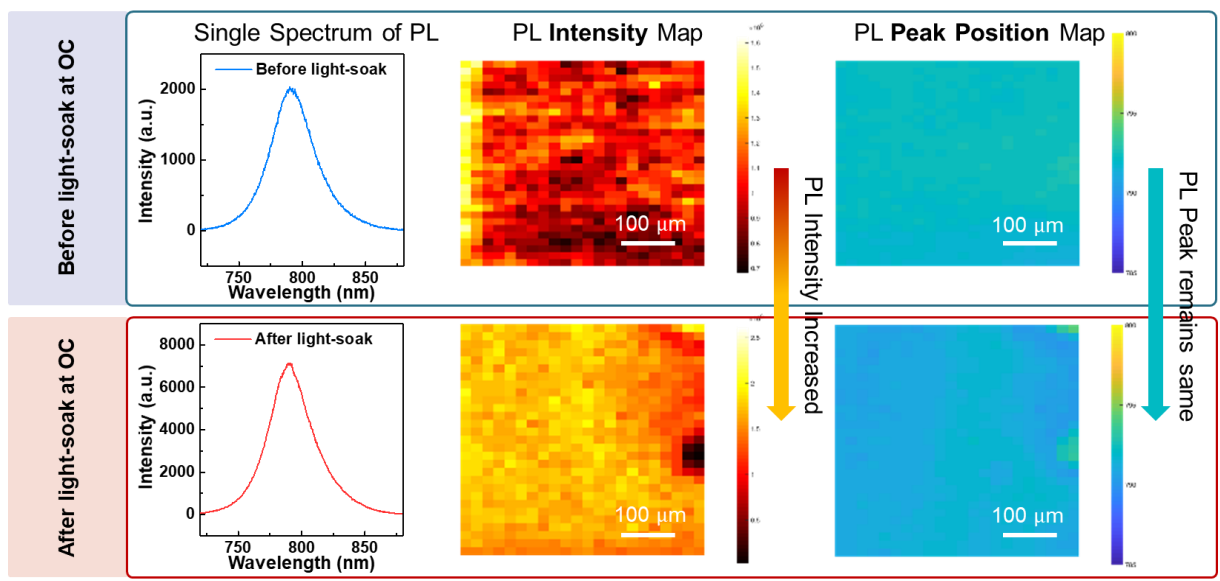

**Fig. S9.** The micro-photoluminescence (μ-PL) mapping for the perovskite device before and after light-soaking at OC condition.

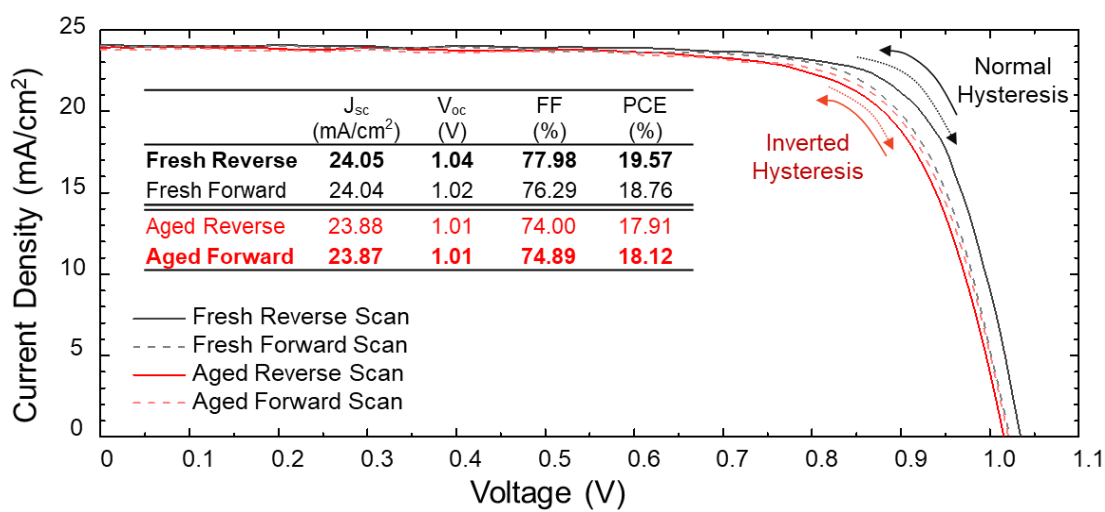

**Fig. S10. Current density curves with reverse scan (from $V_{oc}$ to $J_{sc}$) and forward scan (from $J_{sc}$ to $V_{oc}$) for the PSCs before aging and after aging.** Inverted hysteresis occurs after the aging, while normal hysteresis occurs before the aging. The reverse scan and the forward scan of fresh device (Black line and black dotted line) and device after 16 h of aging under light illumination and electrical forward bias (red line and red dotted line).

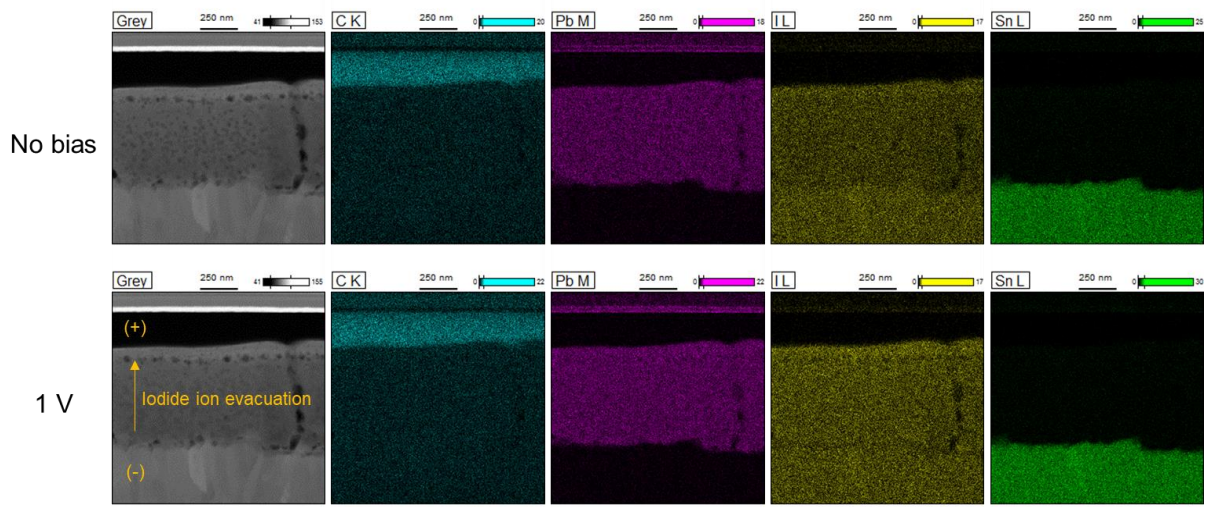

**Fig. S11.** EDS mapping for the nano solar cell before and after applying electrical bias of 1V. The distributions of C, Pb, I and Sn elements are mapped.

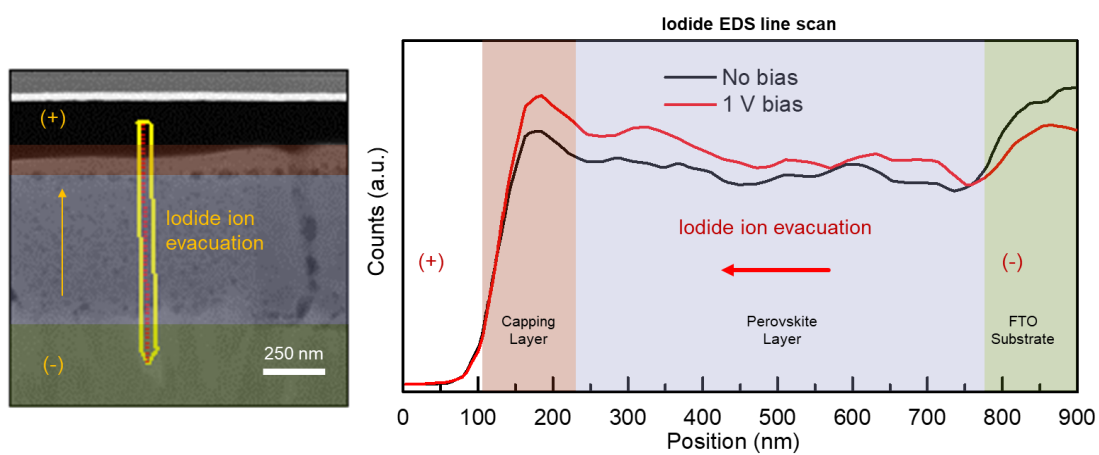

**Fig. S12.** EDS line scan of iodide element for nano solar cell before and after applying electrical bias of 1V.

| (a) OC (Open-circuit) | $J_{sc}$(mA/cm$^2$) | $V_{oc}$(V) | FF (%) | PCE(%) |
|---|---|---|---|---|
| **Initial** | 23.6 | 1.04 | 76.2 | 18.7 |
| **Aged** | 23.5 | 0.93 | 65.4 | 14.3 |
| **Dark recovered (16 h)** | 23.9 | 0.98 | 80.3 | 18.8 |

| (b) SC (Short-circuit) | $J_{sc}$(mA/cm$^2$) | $V_{oc}$(V) | FF (%) | PCE(%) |
|---|---|---|---|---|
| **Initial** | 23.8 | 1.07 | 71.5 | 18.2 |
| **Aged** | 23.6 | 1.06 | 64.0 | 16 |
| **Dark recovered (16 h)** | 22.9 | 1.07 | 75.5 | 18.5 |

| (c) Thermal Recovery | $J_{sc}$(mA/cm$^2$) | $V_{oc}$(V) | FF (%) | PCE(%) |
|---|---|---|---|---|
| **Initial** | 23.9 | 1.03 | 75.6 | 18.6 |
| **Aged** | 23.1 | 0.91 | 63.7 | 13.4 |
| **Thermal recovered (3h)** | 24.0 | 1.03 | 71.5 | 17.8 |

**Table S1.** Photovoltaic parameters for perovskite solar cells for the initial state, after 16 h of aging, and after 16 h of recovery at dark state at (a) open-circuit condition and (b) short-circuit condition. (c) Photovoltaic parameters for perovskite solar cells for the initial state, after 16 h of aging at open-circuit condition, and after 3 h of recovery at dark state and 50 °C.